\documentstyle[12pt]{article}
\def\ba{\begin{eqnarray}}
\def\ea{\end{eqnarray}}

\def\lb{\label}
\def\be{\begin{equation}}
\def\ee{\end{equation}}

\begin{document}

\title{Heterotic geometry without isometries}

\author{A. P. Isaev and O. P. Santillan \thanks{Bogoliubov Laboratory of Theoretical Physics,
JINR, 141 980 Dubna, Moscow Reg., Russia; isaevap@thsun1.jinr.ru,
firenzecita@hotmail.com and osvaldo@thsun1.jinr.ru}}
\date {}
\maketitle

\begin{abstract}

     We present some properties of hyperkahler torsion
(or heterotic) geometry in four dimensions that make it even more
tractable than its hyperkahler counterpart. We show that in $d=4$
hypercomplex structures and weak torsion hyperkahler geometries
are the same. We present two equivalent formalisms describing such
spaces, they are stated in the propositions of section 1. The
first is reduced to solve a non-linear system for a doublet of
potential functions, first found by Plebanski and Finley.
The second is equivalent to finding the solutions of a quadratic
Ashtekar-Jacobson-Smolin like system, but without a volume
preserving condition. This is why heterotic spaces are simpler
than usual hyperkahler ones. We also analyze the strong version of
this geometry. Certain examples are presented, some of them are
metrics of the Callan-Harvey-Strominger type and others are not.
In the conclusion we discuss the benefits and disadvantages of
both formulations in detail.

\end{abstract}

\section{Introduction}

  Supersymmetric $\sigma$ models in two dimensions
with $N=2$ supersymmetry occur on Kahler manifolds while $N=4$
occurs on hyperkahler spaces \cite{Alvarez}. More general supersymmetric
$\sigma$ models can be constructed by including Wess-Zumino-Witten
type couplings in the action \cite{Witten}-\cite{Sierra}. These
couplings can be interpreted as torsion potentials and the
relevant geometry is a generalization of the Kahler and
hyperkahler ones with closed torsion. Such spaces are known as
strong Kahler and hyperkahler torsion (HKT) geometries.
The first examples of N=(4,4) SUSY sigma model with torsion
(and the corresponding HKT geometry) were found in \cite{ivokrivo1}
and developed further in \cite{ivokrivo2}-\cite{ivokrivo4}
by using of the harmonic superspace formalism.

        Hyperkahler torsion geometry is also an usefull mathematical
tool in order to construct
heterotic string models \cite{Sen}-\cite{Hull}. In particular
heterotic $(4,0)$ supersymmetric models are those that lead to
strong hyperkahler torsion geometry
\cite{Bergshoeff}-\cite{Spindel}. If the torsion is not closed
we have a weak hyperkahler geometry and this case has also
physical significance \cite{Callan}-\cite{Gibpapa}.

     A direct way of classifying the possible
HKT spaces is to find the most general weak
hyperkahler spaces and after that to impose the strong condition,
i.e, the closure of the torsion. Callan,
Harvey and Strominger noticed that under a conformal
transformation any usual hyperkahler geometry is mapped into one
with torsion \cite{Callan}; such examples are sometimes called
minimal in the literature \cite{Tod} and they are not the most general
\cite{Valent}. As there is a
particular interest in constructing spaces with at least one
tri-holomorphic Killing vector due to applications related
to dualities \cite{Rocek}, there were found
heterotic extensions of
the known hyperkahler spaces and, in particular, the Eguchi-Hanson
and Taub-Nut ones. The heterotic Eguchi Hanson geometry was shown
to be conformal to the usual one, while heterotic Taub-Nut is a
new geometry \cite{Valent}.

   One of the main properties of hyperkahler torsion spaces is the integrability of the
complex structures, that is,
the annulation of their Nijenhuis tensor. In four dimensions
this implies that the Weyl
tensor of such manifolds is self-dual
\cite{Strachan}-\cite{Finley}. The converse of this statement
is not true in general. On the other hand there exist a one to one
correspondence between four dimensional self-dual structures with
at least one isometry and 3-dimensional Einstein-Weyl structures
\cite{JonTod}. The Einstein-Weyl condition is a generalization of
the Einstein one to include conformal
transformations, and weak hyperkahler spaces should correspond with
certain special Einstein-Weyl metrics. In
\cite{Papadopoulos}-\cite{Tod} it was shown that the self-dual
spaces corresponding to the round three sphere and the Berger
sphere (which are Einstein-Weyl) are of heterotic type.
Arguments related to the
harmonic superspace formalism suggest that indeed there are more
examples \cite{Ivanov}-\cite{Delduc}.

    The present work is related to the construction of weak heterotic geometries
in $d=4$ without Killing vectors. This problem is of interest also
because any weak space with isometries should arise as subcases
of those presented here. For the sake of clarity we resume the result
presented in this letter in the following two equivalent
propositions.
\\

{\bf Proposition 1}{ \it Consider a metric $g$ defined on a
manifold $M$ together with three complex structures $J^{i}$
satisfying the algebra $J^{i}\cdot
J^{j}=-\delta_{ij}+\epsilon_{ijk}J^{k}$ and for which the metric
is quaternion hermitian, i.e, $g(X,Y)=g(J^i X, J^i Y)$. Define the
conformal family of metrics $[g]$ consisting of all the metrics
$g'$ related to $g$ by an arbitrary conformal transformation.
\\

a) Then we have the equivalence \be\lb{equivalencia}
d\overline{J}^i + \alpha \wedge \overline{J}^i=0
 \qquad\Longleftrightarrow\qquad N^i(X,Y)=0,
 \ee
where $\overline{J}^i$ and $N^i(X,Y)$ are the Kahler form and the
Niejenhuis tensor associated to $J^{i}$, $d$ is the usual exterior
derivative and $\alpha$ is a 1-form. If any of (\ref{equivalencia})
hold for $g$, then (\ref{equivalencia}) is also satisfied for any $g'$
of the conformal structure $[g]$, i.e, (\ref{equivalencia}) is
conformally invariant.
\\

 b) Any four dimensional weak hyperkahler torsion metric is equivalent
to one satisfying (\ref{equivalencia}) and there exists a local
coordinate system $(x, y, p, q)$ for which the metric take the form
\be\lb{metropo} g= (dx -\Phi_x dp + \Phi_x dq)\otimes dp +(dy +
\Psi_y dp -\Psi_x dq)\otimes dq, \ee
up to a conformal transformation $g\rightarrow \omega^2 g$. The potentials $\Psi$ and
$\Phi$ satisfy the non-linear system \be\lb{maestro} [\Phi_y
\partial_x
\partial_x + \Psi_x \partial_y \partial_y -
(\Phi_x + \Psi_y)\partial_x \partial_y+\partial_x \partial_p
+\partial_y \partial_q ]\left(%
\begin{array}{c}
  \Phi \\
  \Psi \\
\end{array}%
\right)=0. \ee

c) Conversely any metric (\ref{metropo}) defines a conformal family
$[g]$ in which all the elements $g'$ are weak hyperkahler torsion
metrics. The torsion $T$ corresponding to (\ref{metropo}) is given
by
$$
T=-\Xi_x dq\wedge (dy \wedge dx+\Phi_y dp\wedge dx- \Psi_y
dy\wedge dp)
$$
\be\lb{mono} +\Xi_y dp\wedge (dy\wedge dx +\Phi_x dy\wedge dq
-\Psi_x dq\wedge dx),\ee where $\Xi=\Phi_x-\Psi_y$.  Under the
conformal transformation $g \rightarrow \omega^2 g$ the torsion is
transformed as $ T\rightarrow T + \ast_{g} 2 d\log (\omega)$\footnote{
The action of the Hodge star $\ast_{g}$ is defined by
$$
\ast_{g}e^a=\epsilon_{abcd}e^b \wedge e^c \wedge e^d.
$$}}.
\\

Proposition 1 should not be considered as a generalization of the Kahler formalism for weak
HKT spaces. Although $(\Phi, \Psi)$ is a doublet potential, the metric (\ref{metropo}) is
not written in complex coordinates. The use of holomorphic coordinates for such spaces is
described in detail in \cite{Valent}. The following is an Ashtekar-Jacobson-Smolin like
formulation for the same geometry.
\\

 {\bf Proposition 2}{ \it Consider a
representative $g=\delta_{ab}e^a \otimes e^b$ of a conformal family $[g]$ defined on a
manifold $M$ as at the beginning of Proposition 1, $e^a$ being tetrad 1-forms for which
the metric is diagonal.
\\

 a) Then all the elements $g'$ of $[g]$ will be weak hyperkahler torsion
iff
$$
[e_1,e_2] + [e_3,e_4]= - A_2 e_1 + A_1 e_2 - A_4 e_3 + A_3 e_4
$$
\be\lb{geno} [e_1,e_3]+[e_4,e_2]= - A_3 e_1 + A_4 e_2 + A_1 e_3-
A_2 e_4 \ee
$$
[e_1,e_4]+[e_2,e_3]= - A_4 e_1 - A_3 e_2 + A_2 e_3 + A_1 e_4
$$
where $e_a$ is the dual tetrad of $e^a$ and $A_i$ are arbitrary functions
on $M$.
\\

b) Conversely any solution of (\ref{geno}) defines a conformal family $[g]$ in which all the
elements $g'$ are weak hyperkahler torsion metrics. The torsion $T$ corresponding to
(\ref{metropo}) is given by \be\lb{maza} T=\ast_{g} (A_a-c^b_{ab})e^a \ee where $c^c_{ab}$
are the structure functions defined by the Lie bracket $[e_a,e_b]=c_{ab}^{c}e_c$.
The transformation of the torsion under $g \rightarrow \omega^2 g$ follows directly from
Proposition 1.}
\\

  To conclude, we should mention that the
properties of hyperkahler torsion geometry in higher dimension
were considered for instance, in \cite{Poon1}-\cite{Ofer2}. In
particular, the quotient construction for HKT was achieved in
\cite{Poon3}. Also it was found
that when a sigma model is coupled to gravity the resulting
target metric is a generalization of quaternionic Kahler
geometry including torsion \cite{Ofer}.
This result generalizes the classical one given by Witten
and Bagger \cite{Wite}, who originally did not include a Wess-Zumino
term to the action. A generalization
of the Swann extension for quaternion Kahler torsion spaces was achieved
\cite{Poon2}. To the knowledge of the authors the harmonic superspace
description of quaternion Kahler geometry was already obtained in \cite{Ivi}, \cite{Ogo}
but the extension to the torsion case is still an open problem.

   The present work is organized as follows.
In section 2 we define what is weak and strong heterotic geometry.
We show that the problem to finding weak examples is equivalent to
solving a conformal extension of the hyperkahler condition (namely,
the first (\ref{equivalencia}) given above) together with the
integrability condition for the complex structure. In the third
section we present the consequences of (\ref{equivalencia})
following a work of Plebanski and Finley \cite{Finley}. We
found out that integrability and the conformal extension of the
hyperkahler condition are equivalent in $d=4$, which is the point
a) of Proposition 1. Therefore weak heterotic geometry and
hypercomplex structures are exactly the same concept. We also show
that strong representatives of a given heterotic structure are
determined by a conformal factor satisfying an inhomogeneous
Laplace equation. Just in the case when the structure contains an
hyperkahler metric it is possible to eliminate the inhomogeneous part
by a conformal transformation. We discuss our results in the
conclusions, together with possible applications. For completeness, we show in
the appendix how this geometry arise in the
context of supersymmetric sigma models.

\section{Hyperkahler torsion manifolds}

\subsection{Main properties}

   In this section we define what is hyperkahler torsion
geometry (for an explanation of its physical meaning see Appendix
and references). We deal with
$4n$-dimensional Riemannian manifolds $M$ with metric expressed as \be\lb{numeral}
g=\delta_{ab}e^a\otimes e^b. \ee Here $e^a$ is a tetrad basis for which the metric is
diagonal, and it is defined up to an $SO(4n)$ rotation. It is convenient to
introduce the $4n \times 4n$ matrices
$$
J^{1}=\left(\begin{array}{cccc}
  0 & -I_{n \times n} & 0 & 0 \\
  I_{n \times n} & 0 & 0 & 0 \\
  0 & 0 & 0 & -I_{n \times n} \\
  0 & 0 & I_{n \times n} & 0
\end{array}\right),\;\;\;\;
J^{2}=\left(\begin{array}{cccc}
  0 & 0 & -I_{n \times n} & 0 \\
  0 & 0 & 0 & I_{n \times n} \\
   I_{n \times n} & 0 & 0 & 0 \\
  0 & -I_{n \times n} & 0 & 0
\end{array}\right)
$$
\be\lb{reprodui} J^{3}=J^{1}J^{2}=\left(\begin{array}{cccc}
  0 & 0 & 0 & -I_{n \times n} \\
  0 & 0 & -I_{n \times n} & 0 \\
  0 & I_{n \times n} & 0 & 0 \\
  I_{n \times n} & 0 & 0 & 0
\end{array}\right).
\ee
Then it can be immediately be checked that (\ref{reprodui}) satisfy the multiplication rule
of the quaternions \be\lb{almcomp} J^{i}\cdot J^{j}=-\delta_{ij}+\epsilon_{ijk}J^{k}. \ee
In particular from (\ref{almcomp}) it is seen that $J^i\cdot J^i=-I$,
a property that resembles the
condition $i^2=-1$ for the imaginary unity. The matrices (\ref{reprodui})
are of course not the only ones satisfying this property, in fact any simultaneous
$SO(4n)$ rotation of $J^i$ leaves the multiplication (\ref{almcomp}) unchanged.
With the help of (\ref{reprodui}) we define the $(1,1)$ tensors
\be\lb{unaal} J^{i}=(J^{i})^{b}_{\;a} e_b \otimes e^a \ee which are known as almost complex
structures. Here $e_a$ denote the dual of the 1-form $e^a$. Let us introduce the triplet
$\overline{J}^i$ of $(0,2)$ tensors by
\be\lb{susu} \overline{J}^i(X,Y)=g(X, J^i Y).
\ee One can easily see that the metric (\ref{numeral}) is quaternion
hermitian with respect to any of the complex structures (\ref{unaal}), that is
\be\lb{defcon} g(J^i X, Y)=-g(X, J^i Y) \ee for any X,Y in $T_x M$ (in this notation $J^i X$
denotes the contraction of $J^i$ with $X$). By virtue of (\ref{defcon}) the tensors
(\ref{susu}) are skew-symmetric and define locally a triplet of 2-forms which in the
vielbein basis take the form
\be\lb{dosal}
\overline{J}^i=(\overline{J}^i)_{ab} e^a \wedge e^b.
\ee
The 2-forms (\ref{dosal}) are known as the hyperkahler triplet.
\\

\textbf{Definition}
\\

Heterotic geometry (or torsion hyperkahler geometry) is defined by
the following requirements.
\\

(1) \textit{Hypercomplex condition}. The almost complex structures (\ref{sdcom}) should be
integrable. This mean that the Niejenhuis tensor \be\lb{nijui}
N^i(X,Y)=[X,Y]+J^{i}[X,J^{i}Y]+J^{i}[J^{i}X,Y]-[J^{i} X,J^{i} Y] \ee associated with the
complex structure $J^{i}$, is zero for every pair of vector fields $X$ and $Y$ in $TM_x$.
\\

 (2) \textit{Existence of torsion}. There exists a torsion tensor
$T^{\mu}_{\nu\alpha}$ defined in terms of the metric for which the
$T_{\mu\nu\alpha}=g_{\mu\xi}T^{\xi}_{\nu\alpha}$ is fully
skew-symmetric and therefore define a three form (we write it in the tetrad
basis)
\be\lb{tor}
T= \frac{1}{3!}T_{\mu\nu\lambda}dx^{\mu}\wedge dx^{\nu}\wedge
dx^{\lambda}.
\ee
Here greek indices denote the quantities related
to a coordinate basis $x^{\mu}$\footnote{In the vienbein basis
relation (\ref{tor}) will be
$$
T= \frac{1}{3!}T_{abc}e^a\wedge e^b \wedge e^c.
$$}. If (\ref{tor}) is closed, i.e, $dT=0$, the geometry
will be called strong, otherwise it is called weak.
\\

(3) \textit{Covariant constancy of $J^{i}$}.
Let us define a derivative $D_{\mu}$ with the
connection \be\lb{conoc}
\Upsilon^{\rho}_{\mu\nu}=\Gamma^{\rho}_{\mu\nu}-{1\over
2}T^{\rho}_{\mu\nu}, \ee
$\Gamma^{\rho}_{\mu\nu}$ being the
Christofell symbols of the Levi-Civita connection. By definition the
structures $(J^i)_{a}^{b}$ of hyperkahler torsion spaces
satisfy \be\lb{equo} D_{\mu}(J^i)_{\nu}^{\rho}=0 \ee that is,
they are covariantly constant with respect to $D_{\mu}$.
\\

     We are concerned through this work with $d=4$. In this case the explicit
form of the $(1,1)$ tensors (\ref{unaal}) is
$$
J^1 = -e_1 \otimes e^2 + e_2 \otimes e^1 -e_3\otimes e^4 + e_4\otimes e^3
$$
\be\lb{sdcom} J^2 = -e_1 \otimes e^3 + e_3 \otimes e^1 - e_4\otimes e^2 + e_2\otimes e^4 \ee
$$
J^3 = - e_1 \otimes e^4 + e_4 \otimes e^1 - e_2\otimes e^3 + e_3\otimes e^2,
$$
and the action of (\ref{sdcomo}) over the tangent
space $TM_x$ is defined by \be\lb{acshon}
\begin{array}{rclrclrclrcl}
J^1 ( e_1 ) =\ e_2, &J^1( e_2 ) =-e_1, &J^1 ( e_3 ) =\ e_4, &J^1 ( e_4 ) =-e_3,
\\
J^2  ( e_1 ) =\ e_3, &J^2 ( e_2 ) =-e_4, &J^2 ( e_3 ) =-e_1, &J^2 ( e_4 ) =\ e_2,
\\
J^3 ( e_1 ) =\ e_4, &J^3 ( e_2 ) =\ e_3, &J^3 ( e_3 ) =- e_2, &J^3 ( e_4 ) =- e_1.
\end{array}
\ee
The hyperkahler triplet (\ref{dosal}) is given by
$$
\overline{J}^1 = e^2\wedge e^1 + e^4\wedge e^3
$$
\be\lb{sdcomo} \overline{J}^2 = e^3\wedge e^1 + e^2\wedge e^4  \ee
$$
\overline{J}^3 = e^4\wedge e^1 + e^3\wedge e^2,
$$
By a discussion given above (\ref{reprodui}) with $n=1$ are a non unique
$4\times 4$ representation of the
algebra (\ref{almcomp}) but defined up to an $SO(4)$ rotation.
It follows from (\ref{defcon}) that the metric $g(X,Y)$ is
quaternion hermitian with respect to any $SO(4)$ rotated complex structures.

       Condition $N^{i}(X,Y)=0$ for every $J^{i}$ in (\ref{sdcom}) is equivalent to
$$
[e_1,e_2] + [e_3,e_4] = - A_2 e_1 + A_1 e_2 - A_4 e_3 + A_3 e_4
$$
\be\lb{genasht} [e_1,e_3] + [e_4,e_2] = - A_3 e_1 + A_4 e_2 + A_1 e_3 - A_2 e_4 \ee
$$
[e_1,e_4] + [e_2,e_3] = - A_4 e_1 - A_3 e_2 + A_2 e_3 + A_1 e_4
$$
with certain functions $A_i$ whose form depends on the metric in consideration
\cite{Strachan}. Equations (\ref{genasht}) holds by simply evaluating $N^i(e_a, e_b)=0$
with the
use of (\ref{nijui}) and (\ref{acshon}), the calculation is straightforward and we omit it.
The system (\ref{genasht}) is invariant under a transformation
$$
e_a\rightarrow \omega e_a \; (g\rightarrow \omega^2 g),\;\;\;\;\; A_a\rightarrow A_a + e_a \log \omega.
$$
Therefore the integrability condition is
conformal invariant, that is, any metric $g$ with integrable complex structures define a
conformal structure $[g]$ with the same property. The family $[g]$ is called hypercomplex
structure. Hypercomplex condition implies, but is not implied by, that the Weyl tensor of
$[g]$ is self-dual \cite{Boyer}.  In the limit $A_i=0$ (\ref{genasht}) reduces to a system
equivalent to the Ashtekar-Jacobson-Smolin one \cite{Ashtekar} for hyperkahler spaces.
Nevertheless one can not conclude that (\ref{genasht}) always describes spaces which are
conformally equivalent to hyperkahler ones, even in the case $A_a=0$. \footnote{In order to
obtain an hyperkahler manifold the tetrad should also preserve certain volume form. If this
condition holds the structures are called minimal.} In the special cases in which the
resulting family is conformal hyperkahler, we will say that is of the Callan-Harvey-Strominger
type \cite{Callan}.

   It is important to recall that although the complex structures
are defined up to certain $SO(4)$ automorphisms of
(\ref{almcomp}), this does not affect the integrability condition $N^{i}(X,Y)=0$.
A $SO(4)$ rotation of the complex structures can be compensated
by an $SO(4)$ rotation of the frame $e^a$ leaving (\ref{acshon}) and therefore
$N^a(e_i,e_j)=0$ invariant. By the results to be presented below it
will be clear that such rotations also do not affect conditions 2 and 3.

\subsection{Relation with the Plebanski-Finley conformal structures}

   Requirements 1, 2 and 3 of the definition of HKT geometry are not independent
because the last two imply the first. To see why is it so, let us write
the expression for the Nijenhuis
tensor in a coordinate basis $x^{\mu}$, \be\lb{lacord} N^{\rho}_{\mu\nu}=
(J^i)_{\mu}^{\lambda}[\partial_{\lambda} (J^i)_{\nu}^{\rho}\ -\
\partial_{\nu} (J^i)_{\lambda}^{\rho}]\ -(\mu\ \leftrightarrow \
\nu)=0. \ee Then (\ref{lacord}) together with (\ref{equo}) implies
that \be\lb{coord} T_{\rho\mu\nu}
-(J^i)_{[\mu}^{\lambda}(J^i)_{\nu}^{\sigma}T_{\rho]\lambda\sigma}=0.
\ee
It is simple to check that (\ref{coord}) is satisfied for any skew-symmetric
torsion. Let us consider the tetrad basis $e^a$ for which the complex structures
take the form (\ref{reprodui}), then (\ref{coord}) is a direct
consequence of the skew symmetry of $T_{\mu\nu\alpha}$. Since $N^i(X,Y)$ is a
tensor, if is zero in one basis
then it is zero in any basis. Thus we have proved that the requirements 2 and 3 imply
first. We will reach to the same conclusion in the next section.

    We will show now that the task to find a weak HKT geometry
is in fact equivalent to obtaining the solutions of
\be\lb{miruta}
d\overline{J}^i + \alpha \wedge \overline{J}^i=0
\ee
where $\alpha$ is an arbitrary 1-form. To prove this
we note that the relation
$$
2D_{\mu} g_{\nu\alpha}=- g_{\mu\xi}T^{\xi}_{\nu\alpha} -
g_{\nu\xi}T^{\xi}_{\mu\alpha}
$$
together with the skew symmetry of
$T_{\mu\nu\alpha}=g_{\mu\xi}T^{\xi}_{\nu\alpha}$ yields directly
that \be\lb{yeah} D_{\mu} g_{\nu\alpha}=0. \ee Equation
(\ref{yeah}) implies the equivalence \be\lb{poh}
D_{\mu}(J^i)^{\nu}_{\alpha}=0\qquad \Longleftrightarrow
\qquad D_{\mu}(\overline{J}^i)_{\nu\alpha}=0,\ee and therefore, for
any hyperkahler torsion space
$$
D_{\mu}(\overline{J}^i)_{\nu\alpha}=0.
$$
From the identity \cite{Futa}
$$
D_{\mu}(\overline{J}^i)_{\nu\alpha}=D_{[\mu}(\overline{J}^i)_{\nu\alpha]}-3
D_{[\mu}(\overline{J}^i)_{\xi\rho]}(J^i)^{\xi}_{\nu}
(J^i)^{\rho}_{\alpha} + (\overline{J}^i)_{\mu\xi}
(N^i)^{\xi}_{\nu\alpha} ,
$$
it is easily seen that \be\lb{cuba}
D_{\mu}(\overline{J}^i)_{\nu\alpha}=0\qquad\Longleftrightarrow\qquad
D_{[\mu}(\overline{J}^i)_{\nu\alpha]}=0,\qquad N^i(X,Y)=0.
\ee
where the brackets denote the totally antisymmetric combination of indices.
However we have seen that $N^i(X,Y)=0$ is satisfied by our hypothesis.
A direct calculation using (\ref{equo}),(\ref{conoc})  and the skew-symmetry
of $T_{\mu\nu\alpha}$ shows that \be\lb{hap}
D_{[\mu}(\overline{J}^i)_{\nu\alpha]}=0\qquad
\Longleftrightarrow\qquad d\overline{J}^i +{1\over
2}T_{dbc}(J^i)_{a}^{d}e^{a}\wedge e^{b}\wedge e^{c}=0. \ee The
second (\ref{hap}) is the generalization of the hyperkahler
condition for torsion manifolds. If the torsion
$T_{\mu\nu\alpha}$ is zero, then the second (\ref{hap}) implies
that the hyperkahler triplet is closed, which is a well known
feature of hyperkahler geometry. Consider now the explicit form of
(\ref{hap}) for $\overline{J}^1$. Then (\ref{equo}) and the first
(\ref{reprodui}) gives us that
$$
d\overline{J}^1+T_{234}e^1\wedge e^3 \wedge e^4-T_{134}e^2\wedge e^3 \wedge e^4
+T_{312}e^4\wedge e^1 \wedge e^2-T_{412}e^3\wedge e^1 \wedge e^2
$$
$$
 =d\overline{J}^1+(T_{234}e^1+T_{134}e^2-T_{123}e^4+T_{124}e^3)\wedge (e^1\wedge
e^2+e^3\wedge e^4)=0
$$
and therefore
$$
d\overline{J}^1+(\ast_g T)\wedge \overline{J}^1=0.
$$
The same is obtained for $\overline{J}^2$ and $\overline{J}^3$, namely
\be\lb{ejem} d\overline{J}^i + (\ast_g T)\wedge \overline{J}^i=0. \ee
Let us define now the
1-form $\alpha$ given by \be\lb{alfo} \alpha=\ast_{g} T \qquad \Longleftrightarrow \qquad \ast_{g}
\alpha=T. \ee Then we have from (\ref{ejem}) that (\ref{hap}) is equivalent to \be\lb{pito}
d\overline{J}^i +\alpha \wedge \overline{J}^i=0. \ee
Therefore formulae (\ref{poh})-(\ref{hap}) implies that hyperkahler torsion
geometry is completely characterized by (\ref{miruta}) which is what we
wanted to show.

  Condition (\ref{pito}) is the conformally invariant extension of the
hyperkahler one and was studied by Plebanski and Finley in
the seventies \cite{Finley}. A detailed calculation shows that
(\ref{pito}) is invariant under
\be\lb{trofo2}
e^a\rightarrow \omega e^a\;(g\rightarrow \omega^2 g)\qquad
 \alpha\longrightarrow \alpha + 2 d\log
\omega, \ee and, in consequence, it defines a conformal family $[g]$.
The transformation law (\ref{trofo2}) implies in particular that
if $\alpha$ is a gradient there exist a representative of $[g]$
for which
$$
\alpha=0\qquad \Longleftrightarrow \qquad d\overline{J}^i=0.
$$
Clearly such families have an hyperkahler element and are of the
Callan-Harvey-Strominger type. The main result of Plebanski and Finley is that in general
(\ref{pito}) implies that the Weyl tensor of $[g]$ is self-dual \cite{Finley}. From
(\ref{alfo}) we deduce that under $g \rightarrow \omega^2 g$ it follows that \be\lb{trofo3}
T\longrightarrow T + \ast_{g} 2 d \log (\omega), \ee which is a transformation found by
Callan et all in \cite{Callan}.

   Some more comments are in order. As we have seen, requirements 2 and 3
imply that $N^i(X,Y)=0$ and also imply (\ref{pito}). Therefore
\be\lb{imp1}
N^i(X,Y)=0 \qquad \Longleftarrow \qquad d\overline{J}^i + \alpha
\wedge \overline{J}^i=0. \ee
However it is also known the implicature
\cite{Strachan}-\cite{Boyer}
 \be\lb{imp}
N^i(X,Y)=0 \qquad \Longrightarrow \qquad d\overline{J}^i + \alpha
\wedge \overline{J}^i=0. \ee
To see how (\ref{imp}) holds consider
the connection $\omega$ given by the first structure equation
$$
de^a + \omega^a_{b}\wedge e^b = 0.
$$
It is well known that the antisymmetric part $\omega^a_{[bc]}$ is
related to the structure functions defined by the Lie bracket
$[e_a,e_b]=c_{ab}^{c}e_c$ by
$$
\omega^a_{[bc]} = \frac{1}{2} c_{bc}^{a}\,.
$$
Consider now the form
$$
\alpha = A-\chi
$$
where \be\lb{conectar} A = A_a e^a ,\;\;\;\; \chi =c_{ab}^{b} e^a.
\ee and the self-dual form $\overline{J}^1 = e^1 \wedge e^2+ e^3
\wedge e^4$. Then by use of (\ref{genasht}) one obtain that
$$
d \overline{J}^1 = d(e^1 \wedge e^2+ e^3 \wedge e^4)
$$
$$
= -\frac{1}{2} c^{[1}_{ab}  e^{2]} \wedge e^a \wedge e^b
-\frac{1}{2} c^{[3}_{ab} e^{4]} \wedge e^a \wedge e^b
$$
$$
= e^1\wedge e^2\wedge (c_{ab}^{a} e^b + A_3 e^3 + A_4 e^4)
+e^3\wedge e^4\wedge (c_{ab}^{a} e^b + A_1 e^1 + A_2 e^2)
$$
$$
=(e^1 \wedge e^2 + e^3 \wedge e^4) \wedge ( A-\chi)
$$
and therefore
$$
d \overline{J}^1+ \overline{J}^1\wedge (\chi-A)=0.
$$
This is equivalent to (\ref{oso}) with $\alpha=\chi-A$. The same
formula holds for $\overline{J}^2$ and $\overline{J}^3$. This means that we have
shown the equivalence
\be\lb{imp3}
N^i(X,Y)=0 \qquad \Longleftrightarrow \qquad d\overline{J}^i + \alpha
\wedge \overline{J}^i=0,
\ee
that is, hypercomplex structures are \emph{the same} as the conformal structures defined by
(\ref{pito}). This result refers just to $d=4$.

\section{The general solution}

    We have seen in the previous section that to constructing an hyperkahler torsion space
is equivalent to finding a metric which solves the conditions stated in (\ref{miruta}).
This task was solved by Plebanski and Finley and it is stated in the following proposition
\cite{Finley}.
\\

{\bf Proposition 3} {\it For any four dimensional metric $g$ for which
$$
d\overline{J}^i +\alpha \wedge \overline{J}^i=0
$$
hold there exists a local
coordinate system $(x, y, p, q)$ for which $g$ takes the form
\be\lb{metropo2}
g = (dx -\Phi_x dp + \Phi_x dq)\otimes dp +(dy + \Psi_y dp -\Psi_x dq)\otimes dq,
\ee
up to a conformal transformation $g \rightarrow \omega^2 g$. The functions $\Psi$ and
$\Phi$ depend on $(x, y, p, q)$ and satisfy the non linear system \be\lb{maestro2} [\Phi_y
\partial_x
\partial_x + \Psi_x \partial_y \partial_y -
(\Phi_x + \Psi_y)\partial_x \partial_y+\partial_x \partial_p
+\partial_y \partial_q ]\left(%
\begin{array}{c}
  \Phi \\
  \Psi \\
\end{array}%
\right)=0.
\ee
The converse of this assertion is also true.}
\\

For completeness we give the proof of Proposition 3 in some detail.
It is convenient to
introduce the Penrose notation and write the metric $g$ in the
following complex form \be\lb{compo} g=\delta_{ab}e^a\otimes
e^b=E^1 \otimes E^2 + E^3 \otimes E^4. \ee We have defined the new
complex tetrad
$$
E^1=\frac{1}{\sqrt{2}}(e^1 + ie^2),\qquad
E^2=\frac{1}{\sqrt{2}}(e^1 - ie^2)
$$
\be\lb{penu} E^3=\frac{1}{\sqrt{2}}(e^3 + ie^4),\qquad
E^4=\frac{1}{\sqrt{2}}(e^3 - ie^4).
\ee
In terms of this basis the
complex structures (\ref{sdcomo}) are expressed as
$$
\overline{J}^1=2 E^4\wedge E^1,\qquad \overline{J}^3=2 E^3\wedge
E^2
$$
\be\lb{penstru} \overline{J}^2=-E^1\wedge E^2 + E^3\wedge E^4. \ee
The advantage of this notation is that allows us to directly apply
the Frobenius theorem.
\\

{\bf Frobenius theorem} {\it In $n$-dimensions, if in a certain
domain $U$ there exist $r$ 1-forms $\beta^i$ ($i=1,..,r$) such that
$$
\Omega=\beta^1\wedge...\wedge \beta^r \neq 0
$$
and there exists a 1-form $\gamma$ such that \be\lb{frobo}
d\Omega=\gamma\wedge\Omega, \ee then there exist some functions
$f^i_j$ and $g^j$ on $U$ such that \be\lb{efe} \beta^i = \sum_{j=1}^r f^i_j dg^j.
\ee}

   The two conditions
\be\lb{oso} d\overline{J}^1 +\alpha \wedge \overline{J}^1=0,
\qquad d\overline{J}^3 +\alpha \wedge \overline{J}^3=0 \ee of
(\ref{pito}) are of the form (\ref{frobo}) with $E^i$ playing
the role of the 1-forms $\beta^i$, $\overline{J}^1$ and
$\overline{J}^3$ are the analogs of $\Omega$ and $\alpha$ play the
role of $\gamma$. Then Frobenius theorem implies the existence of
scalar functions $\widetilde{A},...,\widetilde{H}$ and $p,..,s$ such that
$$
E^1=\widetilde{A} dp + \widetilde{B} dq,
$$
$$
E^2=\widetilde{E} dr + \widetilde{F} ds,
$$
\be\lb{anlo}
E^3=-\widetilde{G} dr - \widetilde{H} ds
\ee
$$
E^4=-\widetilde{C} dp - \widetilde{D} dq.
$$
The functions $\widetilde{A},...,\widetilde{H}$ and  $p,..,s$ are
the analogs of $f^i_j$ and $g^j$ in (\ref{efe}), respectively.

    Consider the two functions $\phi$ and $f$ defined by
$$
\widetilde{A}\widetilde{D}-\widetilde{B}\widetilde{C}=
(\phi^{-1}e^{f})^2,
$$
\be\lb{def} \widetilde{E}\widetilde{H}-\widetilde{F}\widetilde{G}=
(\phi^{-1}e^{-f})^2.
\ee
It is more natural to use the variables
$$
(A,B,C,D)=(\phi^{-1}e^{f})^{-1}(\widetilde{A}, \widetilde{B}, \widetilde{C}, \widetilde{D})
$$
$$
(E,F,G,H)=(\phi^{-1}e^{-f})^{-1}(\widetilde{E}, \widetilde{F}, \widetilde{G}, \widetilde{H})
$$
with the functions $A,..., H$ normalized as
\be\lb{dinamic} AD-BC=1,\qquad EH-GF=1.
\ee
Using (\ref{def}) we can express
(\ref{anlo}) as
$$
E^1=\phi^{-1}e^{f}(A dp + B dq) ,
$$
$$
E^2=\phi^{-1}e^{-f}(E dr + Fds),
$$
\be\lb{anl} E^3=-\phi^{-1}e^{-f}(G dr + H ds) ,\ee $$
 E^4=
-\phi^{-1}e^{f}(C dp + D dq),
$$
Since
$$
E^1\wedge E^2\wedge E^3\wedge E^4=\phi^{-4} dp \wedge dq \wedge dr
\wedge ds \neq 0
$$
one can consider the functions $p$, $q$, $r$ and $s$ as
independent coordinates.

    From (\ref{anl}), (\ref{penstru}) and (\ref{oso}) it follows the
expression for $\alpha$ \be\lb{plin} \alpha=d \log (\phi)+ f_p dp + f_q dq -
f_r dr - f_s ds. \ee
We now consider the remaining condition
(\ref{pito}), namely \be\lb{mm} d\overline{J}^2 +\alpha \wedge
\overline{J}^2=0. \ee By the use of (\ref{penstru}) together with
(\ref{anl}) we find that
$$
\overline{J}^2=
\phi^{-4}[(AE+CG)dp\wedge dr+(AF+CH)dp\wedge ds +(BE+DG) dq\wedge
dr
$$
\be\lb{difo}
+(BF+DH) dq \wedge ds]. \ee A direct calculation
shows that (\ref{mm}) together with (\ref{plin}) implies
$$
(AE+CG)2f_s-(AF+CH)2f_r=(AE+CG)_s-(AF+CH)_r,
$$
$$
(BE+DG)2f_s-(BF+DH)2f_r=(BE+DG)_s-(BF+DH)_r, $$
\be\lb{system}
(AE+CG)2f_q-(BE+DG)2f_p=-(AE+CG)_q + (BE+DG)_p,\ee
$$
(AF+CH)2f_q-(BF+DH)2f_p=-(AF+CH)_q+(BF+DH)_p.
$$
The first two (\ref{system}) show that there exist certain
functions $x$ and $y$ such that
$$
AE+CG=e^{2f}x_r,
$$
$$
BE+DG=e^{2f}y_r,
$$
\be\lb{ping2}
AF+CH=e^{2f}x_s,
\ee
$$
BF+DH=e^{2f}y_s.
$$
By
multiplying the last two (\ref{system}) by $e^{2f}$ and using
(\ref{ping2}) we also obtain that \be\lb{subso}
(e^{4f}x_r)_q=(e^{4f}y_r)_p,\qquad (e^{4f}x_s)_q=(e^{4f}y_s)_p.
\ee
The solution of (\ref{ping2}) and (\ref{subso}) are
$$
G=A(e^{2f}y_r)-B(e^{2f}x_r)
$$
$$
H=A(e^{2f}y_s)-B(e^{2f}x_s)
$$
\be\lb{solo}
E=D(e^{2f}x_r)-C(e^{2f}y_r)
\ee
$$
F=D(e^{2f}x_s)-C(e^{2f}y_s)
$$
The normalization (\ref{dinamic}) and (\ref{solo}) implies that
$$
J=\left|%
\begin{array}{cc}
  x_r &  x_s \\
  y_r &  y_s \\
\end{array}%
\right|=e^{-4f}\neq 0,
$$
and therefore $x$ and $y$ can be used as independent coordinates
instead of $r$ and $s$. Moreover, the functions $A,..,D$ are not
determined by the equations but are just constrained by the first
(\ref{dinamic}). This reflects that the tetrad and the vielbein are
defined up to an $SO(4)$ rotation which leaves the
condition $AD-BC=1$ invariant. Therefore we can
select $A=D=1$ and $B=C=0$ withouth losing generality. The
tetrad becames simplified as
$$
E^1=(\phi e^{-f})^{-1}dp,
$$
$$
E^2=(\phi e^{-f})^{-1}(dx + K dp + L dq),
$$
\be\lb{ij} E^3=-(\phi e^{-f})^{-1}(dy + M dp + N dq), \ee
$$
E^4=-(\phi e^{-f})^{-1}dq,
$$
being $K=-x_p$, $L=-x_q$, $M=-y_p$ and $N=-y_q$. The dual basis of
(\ref{ij}) is
$$
E_2=\phi e^{-f}
\partial_x
$$
$$
E_1=\phi e^{-f}(\partial_p - K \partial_x - M \partial_y) ,
$$
\be\lb{duo} E_4=-\phi e^{-f}(\partial_q - L
\partial_x - N\partial_y), \ee
$$
E_3=-\phi e^{-f}\partial_y.
$$
Since our goal is to find a conformal structure,
it is convenient to eliminate the factor $\phi e^{-f}$ in (\ref{duo}) by a conformal
transformation. Then from (\ref{oso}) it is calculated that \be\lb{fingo}
2\alpha=(L-M)_x dq-(L-M)_y dp. \ee Also, it follows from (\ref{mm}) and
(\ref{fingo}) that
$$
\partial_x K +\partial_y L=0, \qquad \partial_x M +\partial_y  N=0,
$$
\be\lb{cuento}  (\partial_p - K \partial_x - M \partial_y) L
-(\partial_q - L
\partial_x - N\partial_y) K =0,\ee
$$
 (\partial_p - K \partial_x - M \partial_y)
N-(\partial_q - L
\partial_x - N\partial_y) M =0
$$
The first two (\ref{cuento}) imply the existence of two
functions $\Phi$ and $\Psi$ such that \be\lb{pico} K=-\Phi_x,
\qquad L=\Phi_x, \qquad M=\Psi_y, \qquad N=-\Psi_x, \ee and the
last two (\ref{cuento}) together with (\ref{pico}) give the
following non-linear equations for the doublet $(\Phi, \Psi)$
\be\lb{master} [\Phi_y
\partial_x
\partial_x + \Psi_x \partial_y \partial_y -
(\Phi_x+\Psi_y)\partial_x \partial_y+\partial_x \partial_p
+\partial_y \partial_q ]\left(%
\begin{array}{c}
  \Phi \\
  \Psi \\
\end{array}%
\right)=0. \ee
We conclude that the most general structure
satisfying (\ref{pito}) is given in terms of two key functions
$\Phi$ and $\Psi$ satisfying (\ref{master}). A simple calculation shows that
the metric corresponding to the tetrad (\ref{duo}) is (\ref{metropo2}), which is
what we wanted to prove (Q.E.D).
\\

  It will be instructive if we show that for any structure of the proposition 3
it follows that
$$
N^i(X,Y)=0,
$$
in accordance with (\ref{imp1}). This is completely equivalent to solve the
system (\ref{genasht}), that is
$$
[e_1,e_2]+[e_3,e_4]= - A_2 e_1 + A_1 e_2 - A_4 e_3 + A_3 e_4,
$$
$$
[e_1,e_3]+[e_4,e_2]= - A_3 e_1 + A_4 e_2 + A_1e_3- A_2 e_4,
$$
$$
[e_1,e_4]+[e_2,e_3]= - A_4 e_1 - A_3 e_2 + A_2 e_3 + A_1 e_4,
$$
for the basis $e_i$ corresponding to (\ref{metropo2}). From
(\ref{penu}), (\ref{ij}) and (\ref{pico}) we find that
$$
e^1=\frac{1}{\sqrt{2}}[dp +(dx -\Phi_y dp + \Phi_x dq)],
$$
$$
e^2=\frac{1}{i\sqrt{2}}[dp -  (dx - \Phi_y dp + \Phi_x dq)]
$$
\be\lb{penul} e^3=-\frac{1}{\sqrt{2}}[dq + (dy + \Psi_y dp -\Psi_x
dq)],\ee
$$
e^4=\frac{1}{i\sqrt{2}}[dq -(dy + \Psi_y dp -\Psi_x dq)].
$$
The dual basis corresponding to (\ref{penul}) is
$$
e_1=\frac{1}{\sqrt{2}}[(\partial_p +\Phi_y \partial_x - \Psi_y
\partial_y) + \partial_x],
$$
$$
e_2=-\frac{1}{i\sqrt{2}}[(\partial_p +\Phi_y
\partial_x - \Psi_y \partial_y) -  \partial_x]
$$
\be\lb{penulo} e_3=-\frac{1}{\sqrt{2}}[\partial_y +  (\partial_q -
\Phi_x\partial_x +\Psi_x\partial_y)], \ee
$$
e_4=\frac{1}{i\sqrt{2}}[\partial_y -  (\partial_q - \Phi_x
\partial_x +\Psi_x\partial_y)].
$$
 It is simple to check that the first (\ref{genasht}) for
(\ref{penulo}) imply that $A_i=0$. After some tedious
calculation one obtain from the full system (\ref{genasht}) that
$$ [\Phi_y
\partial_x
\partial_x + \Psi_x \partial_y \partial_y -
(\Phi_x+\Psi_y)\partial_x \partial_y+\partial_x \partial_p
+\partial_y \partial_q ]\left(%
\begin{array}{c}
  \Phi \\
  \Psi \\
\end{array}%
\right)=0,
$$
which is (\ref{master}). \emph{This again shows that the
equivalence (\ref{imp3}) is true.}

     We conclude that the general form of an HKT metric is
$$
g= (dx -\Phi_y dp +
\Phi_x dq)\otimes dp +(dy + \Psi_y dp -\Psi_x dq)\otimes dq
$$
up to a conformal transformation.  The form $\alpha$ in (\ref{fingo})
can be expressed as \be\lb{cho} \alpha=\Xi_x E^1-\Xi_y E^4, \ee
where $\Xi=\Phi_x-\Psi_y$. One can prove by using
(\ref{penu}) that \be\lb{usefull} \ast E^1=-E^2\wedge E^3\wedge
E^4,\qquad \ast E^4=-E^1\wedge E^2\wedge E^3. \ee Formula
(\ref{usefull}) together with (\ref{alfo}) and (\ref{cho}) gives
explicitly the torsion, namely \be\lb{takea} T=-\Xi_x E^2\wedge E^3\wedge E^4+\Xi_y
E^1\wedge E^2\wedge E^3. \ee  All the results of this section
are stated in the propositions of the introduction. It is instructive at this point
to consider some simple examples of
hyperkahler torsion spaces.
\\

\textit{Example1} Let us construct weak heterotic
spaces with two commuting Killing vectors. The coordinates will be
named $(\rho, \eta, \theta, \varphi)$. Consider four functions
$f_1,...,f_4$ and $g_1,...,g_4$ depending on the coordinates
$x^1=\rho$ and $x^2=\eta$ and the vector fields
$$
e_j= f_j \partial_{\theta} + g_j \partial_{\varphi}+
\partial_{x^j}, \qquad (j=1,2)
$$
$$
e_j= f_j \partial_{\theta} + g_j
\partial_{\varphi}. \qquad(j=3,4)
$$
The basis $e_j$ is the most general for a metric with two
commuting isometries up to a conformal change. Introducing this
expressions into (\ref{genasht}) gives $A_i=0$ together with the
system
$$
(f_3)_{\eta}-(f_4)_{\rho}=0,\;\;\;\;\;(g_3)_{\eta}-(g_4)_{\rho}=0
$$
\be\lb{sistem}
(f_2)_{\eta}+(f_1)_{\rho}=0,\;\;\;\;\;(g_2)_{\eta}+(g_1)_{\rho}=0
\ee
$$
(f_1)_{\eta}-(f_2)_{\rho}=0,\;\;\;\;\;(g_1)_{\eta}-(g_2)_{\rho}=0
$$
and from the first we see that $f_3=(H)_{\rho}$ and
$f_4=(H)_{\eta}$ for some function $H(\rho, \eta)$. This mean that
we can make the coordinate change $(\rho, \eta, \theta,
\varphi)\rightarrow (\rho-H, \eta, \theta, \varphi)$ and eliminate
$f_3$ and $f_4$. The same holds for $g_3$ and $g_4$. From system
(\ref{sistem}) together with these simplifications we get Cauchy
Riemann equations implying that $f(z)=e_1 - ie_2$ and $g(z)=f_1 -
i f_2$ are holomorphic functions depending on the argument
$z=\rho+i\eta$. The corresponding weak heterotic metric is
\be\lb{simplo} \widetilde{g}=d\rho^2+ d\eta^2 +\frac{(e_1 d\theta
-  f_1 d\varphi)^2 + ( e_2 d\theta - f_2 d\varphi)^2}{(e_{1}
f_{2}-e_{2}f_{1})^2}. \ee The commuting Killing vectors are
$\partial_{\theta}$ and $\partial_{\varphi}$.
\\

\textit{Example 2} Another family of hypercomplex structures
that can be constructed in terms of holomorphic functions. It can directly be checked
that the system
(\ref{genasht}) with the coefficients $A_i=0$ can be cast in the following complex form \be\lb{compash}
[e_1+ i e_2, e_1- i e_2]-[e_3+ i e_4, e_3- i e_4]=0,\;\;\; [e_1+ i e_2, e_3- i e_4]=0 \ee
Let $M$ be a complex surface with holomorphic coordinates $(z_1, z_2)$ and let us define
four vector fields $e_i$ as
$$
e_1 + i e_2=f_1\frac{\partial}{\partial z_1} +
f_2\frac{\partial}{\partial z_2}
$$
$$
e_3 + i e_4=f_3\frac{\partial}{\partial z_1} +
f_4\frac{\partial}{\partial z_2}
$$
being $f_j$ a complex function on $M$. Then (\ref{compash})
implies that $\partial f_j/\partial \overline{z}_k=0$ and
therefore we can construct a family of hypercomplex structures by using four
arbitrary holomorphic functions or two holomorphic vector fields
\cite{Joyce}.
\\

\textit{Example 3} If we set $\Xi=\Phi_x-\Psi_y=0$ in (\ref{cho}) it follows that there
exists a function $\Theta$ such that
$$
\Phi=\Theta_y,\qquad \Psi=\Theta_x.
$$
Then (\ref{master}) implies that \be\lb{secod}
\Theta_{xx}\Theta_{yy}-(\Theta_{xy})^2 +
\Theta_{xp}+\Theta_{yq}=0. \ee Equation (\ref{secod}) is known as
the second heavenly equation and the corresponding metric
\be\lb{pleb2} g=dp\otimes(dx-\Theta_{yy}dp+\Theta_{xy}dq)+
dq\otimes(dy+\Theta_{xy}dp-\Theta_{xx}dq) \ee is the most general
hyperkahler one \cite{Plebanski}. There exists a coordinate system
$(r, s, p, q)$ defined in terms of $(x, y, p, q)$ for which the
metric (\ref{pleb2}) became \be\lb{bijo}
g = K_{pr}dp\otimes dr + K_{ps}dp\otimes
ds + K_{qr}dq\otimes dr+K_{qs}dq\otimes ds. \ee The function $K$
plays the role of a Kahler potential and satisfies the first
heavenly equation \be\lb{pleb1}
\left|%
\begin{array}{cc}
  K_{pr} &  K_{ps}\\
 K_{qr} &  K_{qs} \\
\end{array}%
\right|=1. \ee
It is clear that the weak families corresponding to (\ref{pleb1}) or
(\ref{pleb2}) are those of the Callan et.all type. Particular solutions of (\ref{master})
have been found for instance in \cite{Mason}, but the general solution is unknown even
for (\ref{pleb1}).
\\

The geometries presented till now are weak heterotic.
Due to conformal invariance all the elements of the conformal
family $[g]$ corresponding to (\ref{metropo2}) are also weak.
Therefore in order to find an strong geometry we should not impose
$dT=0$ for the torsion (\ref{mono}) corresponding to (\ref{metropo2})
but instead, we must solve $d\widetilde{T}=0$ where
$\widetilde{T}= T + \ast_{g} 2 d \log (\omega)$ is the torsion
corresponding to the representative $\widetilde{g}=\omega^2 g$.
After some calculation we obtain the following \emph{linear}
equation defining $\omega$ \be\lb{laplo} \Delta_{g}
log(\omega)=\partial_{[1}T_{234]}, \ee where $\Delta_{g}$ is the
Laplace operator for (\ref{metropo2}). The left hand side (\ref{laplo})
does not depend on $\omega$ and acts as an inhomogeneous source.
Then the spaces $\widetilde{g}=\omega^2 g$  will be a strong
representatives of the family if $\omega$ solve (\ref{laplo}). The
spaces (\ref{metropo2}) will be strong only in the case that constant
$\omega$ is a solution of the resulting equations, which implies
that \be\lb{cono}
\partial_{[1}T_{234]}=\Xi_{xp}-(\Xi_x \Phi_y)_y+(\Xi_x \Psi_y)_x +
 \Xi_{yq} -(\Xi_y \Phi_x)_x + (\Xi_y \Psi_x)_y=0.
\ee It also follows from the result of this section that if we
deal with a minimal weak structure, then (\ref{laplo}) is reduced
to \be\lb{Calas} \Delta_{\widehat{g}} log(\omega)=0, \ee being
$\widehat{g}$ the hyperkahler representative of the family
\cite{Callan}.

\section{Conclusions}

   In the present work we attacked the problem of constructing
weak and strong 4-dimensional hyperkahler torsion geometries
without isometries. We have shown that the most general local form
for a weak metric is defined by two potential functions satisfying a
non linear system, up to a conformal transformation.

   We argue that the present work has practical applications,
although the system of equations for the potentials is highly
non-linear and difficult to solve. We showed that in four
dimensions hypercomplex structures and weak torsion hyperkahler
geometries are totally equivalent concepts. Concretely speaking,
we showed that integrability of the complex structures implies
and is implied by the other weak HKT properties. Therefore the
problem to find weak heterotic geometries is reduced to find the
solutions of an Ashtekar-Jacobson-Smolin like system without the
volume preserving condition, which has the advantage to be
quadratic.

   From this discussion and the Ashtekar formulation of
hyperkahler spaces it is clear that a
weak HKT metric will be of the Callan-Harvey-Strominger type (i.e.
conformal to an hyperkahler one) if there is a volume form
preserved by its tetrad. Nevertheless not all the
hypercomplex structures preserves a 4-form, the metrics
given in \cite{Valent}-\cite{Tod} are counterexamples. We
have seen also that a space is of Callan-Harvey-Strominger type if
the torsion is the dual of a gradient. Therefore this condition
and the volume preserving one should be equivalent.

   The strong condition (i.e. the closure of the torsion form)
is not invariant under a general conformal transformation and thus
it does not define a conformal structure.
Nevertheless the problem to find the strong representatives of a
weak heterotic family $[g]$ is reduced to solving a \emph{linear}
non homogeneous laplacian equation over an arbitrary element $g$.
The homogeneous part can be eliminated if the metric is
conformal to a hyperkahler one.

   At first sight it looks that the "Ashtekar like" formulation for HKT is
the most convenient formulation. Instead we suggest that the other
variant could be more useful in the context dualities, in which
the presence of isometries is needed \cite{Rocek}. To make this
suggestion more concrete, let us recall that self-dual spaces are
described in terms of a Kahler potential satisfying the heavenly
equation \cite{Plebanski}. Although this equation is non-linear
and the general solution is not known, the analysis of the Killing
equations and the possible isometries allowed to classify the
hyperkahler metrics with one Killing vector \cite{Boyero}. In
particular it was found that they are divided in two types, the
first described for by a monopole equation and the second by some
limit of a Toda system \cite{Gibbons}-\cite{Boyero}. We propose
that it is worthy to generalize, if is possible, the mathematical
machinery to classify the Killing equations of gravitational
instantons \cite{Boyero}-\cite{Przanowski2} to the hyperkahler
torsion case. Perhaps it will be possible to find the
classification of heterotic geometry with one or more Killing
vectors and in particular, to identify the subcases that match for
duality applications. This goal has not been achieved yet, and our
suggestion could be an alternative point of view of the references
\cite{Tod}-\cite{Delduc}.
\\

    We are grateful to G.Valent for orientation
about literature on the subject. O.P.S thanks specially to Luis Masperi for his
permanent and unconditional support during his
mandate at the CLAF (Centro Latinoamericano
de Fisica).
\\

\section{Appendix}

   In this section we sketch why strong torsion hyperkahler geometry is
the target space metric of $N=4$ supersymmetric sigma models \cite{Alvarez}.
Sigma models in dimension two are described by Lagrangians of the
form
 \be\lb{sigma} I_{b}=-\frac{1}{2\pi}\int
d^2x
[g_{AB}(\varphi)\partial_{\mu}\varphi^A\partial^{\mu}\varphi^B +
 b_{AB}(\varphi)\epsilon^{\mu\nu}\partial_{\mu}\varphi^A\partial^{\nu}\varphi^B],
\ee being $g_{AB}$ an arbitrary metric tensor \footnote{In the
main text we have used latin indices (a,b,..) as flat indices and greek
indices ($\alpha$, $\beta$,..) as curved ones. Instead in this appendix
capital latin indices (A,B,..) are curved. We use capital latin
indices and not greek in order to do not confuse them with the greek
ones used to denote space-time derivative $\partial_{\mu}$ or
gamma matrices $\gamma^{\mu}$.}. The fields $\varphi^A$ are
bosonic $(A=1,...,d)$ and depend on the coordinates $x^i$
($i=1,2$). This means that $\varphi^A$ parameterize two-cycles on a
d-dimensional target manifold $M$ with a metric $g_{AB}$. The
second term is a Wess-Zumino-Witten type term.

    If we want to make the supersymmetric
extension of (\ref{sigma}) we need to include a number of fermion degrees equal to the
number of the boson ones. Thus, we need to introduce a set of spinors $\psi'^A$
and impose a supersymmetry transformation, which on general grounds should be of the form
\be\lb{sosos}
\delta\varphi^A=\overline{\varepsilon}_1 J^A_B\psi'^B
\ee
where $J$ is some $(1,1)$ tensor. We
still have the freedom to make the redefinition
$\psi'^B=(J^{-1})^B_A\psi^A$ in (\ref{sosos}), for which
(\ref{sosos}) is converted into
\be\lb{soso}
\delta\varphi^A=\overline{\varepsilon}_1 \psi^A
\ee
The $N=(1,0)$ supersymmetric extension
of (\ref{sigma}) is \cite{Spindel}
\be\lb{action}
 I_{sb}=-\frac{1}{2\pi}\int d^2x
[g_{AB}(\varphi)\partial_{\mu}\varphi^A\partial^{\mu}\varphi^B +
 b_{AB}(\varphi)\epsilon^{\mu\nu}\partial_{\mu}\varphi^A\partial^{\nu}\varphi^B
 + \overline{\psi}^A
\gamma^{\mu}D_{\mu}\psi^B g_{AB}], \ee where the spinors $\psi^A$
are right-handed and $\varepsilon$ is left-handed. The covariant derivative
$D_{\mu}$ is defined by the action
$$
D_{\mu}\psi^A=\partial_{\mu}\psi^A + \Upsilon^A_{BC}\psi^B
\partial_{\mu}\varphi^C,
$$
$$
\Upsilon^A_{BC}=\Gamma^A_{BC}+\frac{1}{2}T^A_{BC}
$$
where $\Gamma^A_{BC}$ are the Christofell symbols of the Levi-Civita
connection for $g_{AB}$. The torsion form $T_{ABC}$ is entirely
defined in terms of the Wess-Zumino-Witten potential $b_{AB}$ as
\be\lb{torso} 2T_{ABC}=-3b_{[AB,C]}\qquad \Longleftrightarrow
\qquad dT=0. \ee The indices $A,B$ are lowered and raised through
$g_{AB}$. Action (\ref{action}) is invariant
under (\ref{soso}) and \be\lb{here} \delta
\psi^A=\gamma^{\mu}\partial_{\mu}\varphi^A
\varepsilon-\psi^B\overline{\varepsilon} S^A_{BC}\psi^C. \ee The
tensor $S_{ABC}$ is constrained by supersymmetry arguments to be
skew symmetric and covariantly constant
$$
S_{ABC;D}=S_{ABC,D}-3S_{E[AB}\Upsilon^E_{C]D}=0.
$$
The presence of a covariantly constant 3-form gives a new
restriction on the manifold. On group manifolds this condition has
solution and $S$ can be chosen proportional to $T$
\cite{Spindel}. The commutator of two supersymmetries on the boson
field is \be\lb{tecnic} [\delta,
\delta']\varphi^A=\overline{\varepsilon}'\gamma^{\mu}\varepsilon
(2\partial_{\mu}\varphi^A
-\overline{\psi}^B\gamma_{\mu}S^A_{BC}\psi^C), \ee and corresponds
to the usual supersymmetry algebra without the presence of central
charges.

      Let us consider now the problem of constructing
an $N=(2,0)$ supersymmetric extension
of (\ref{sigma}). For this purpose we should impose a new supersymmetry transformation to
(\ref{action}) without spoiling (\ref{soso}) and (\ref{here}). As we saw the
general supersymmetry variation for $\varphi^A$ is of the form
\be\lb{ground}
\delta^1\varphi^A=\overline{\varepsilon}_1 (J^{1})^A_B\psi^B, \ee being $J^1$ certain
$(1,1)$ tensor. Due to the invariance of (\ref{action}) with respect to the
symmetry (\ref{here}) we have to impose \be\lb{no} I_{sb}(\varphi,
J^1\psi)=I_{sb}(\varphi,\psi). \ee Formula (\ref{no}) implies that \be\lb{eil}
g_{AB}(J^{1})_{C}^{B} + (J^{1})_{A}^{B}g_{BC} = 0, \ee and then the metric is hermitian
with respect to the tensor $J^{1}$. Condition (\ref{no}) also implies that \be\lb{equon}
D_{A}(J^{1})_{C}^{B}=0\qquad \Longrightarrow \qquad
(J^{1})_{[AB,C]}=-2(J^{1})^{D}_{[A}T_{B,C]D}. \ee This means
that $J^{1}$ is covariantly
constant with respect to the connection with torsion. The compatibility condition of
(\ref{equon}) is \be\lb{integrab} (J^{1})^A_{E}R^E_{BCD}=R^A_{ECD}(J^{1})^E_{B}, \ee and
more generally the tensor $J^{1}$ commute with all the generators of the holonomy group.
The new supersymmetry transformations result
$$
\delta^{1} \varphi^A=\overline{\varepsilon}_{1}(J^{1})^A_B\psi^B.
$$
\be\lb{heru} \delta^1 \psi^A=\gamma^{\mu}\partial_{\mu}\varphi^B
(J^{1})^A_B \varepsilon_{1} -[(J^{1})^A_D \Upsilon^D_{BE}
(J^{1})^E_C + (S_1)^A_{BC}]\psi^B\overline{\varepsilon}_{1}\psi^C.
\ee The tensor $(S_{-})^A_{BC}$ is not determined by other
quantities. The commutator of the two supersymmetries acting over
$\varphi^A$ is
$$
[\delta, \delta^1]\varphi^A=\overline{\varepsilon}_{1}\gamma_{\mu}\varepsilon_{1}
[(J^{1})^A_B + (J^{1})^B_A]\partial_{\mu}\varphi^B
$$
\be\lb{commo} +
\overline{\varepsilon}_{1}\gamma_{\mu}\varepsilon_{1}
\overline{\psi}^B\gamma_{\mu}\psi^C (J^1)^A_{D} N^D_{BC}, \ee plus
terms depending on $(S_{1})^A_{BC}$. Then the usual supersymmetric
algebra holds if (\ref{commo}) is zero and this implies that
\be\lb{nieu} (J^{1})^A_B + (J^{1})^B_A=0,\qquad N_{AB}^C=0,\qquad
(S_{1})^A_{BC}=0. \ee The first equation (\ref{nieu}) together with
(\ref{eil}) implies that $J^{1}$ is an almost complex structure
and therefore the dimension of the target manifold should be even
(d=$2n$). The second (\ref{nieu}) show that $J^{1}$ is integrable.
The metrics $g_{AB}$ for which (\ref{equon}) and (\ref{nieu}) hold
are Kahler torsion. We conclude that $N=2$ supersymmetric sigma
models occur on \emph{Kahler torsion manifolds}.

      If there is a third supersymmetry corresponding to another
complex structure $J^2$ and to a parameter $\varepsilon_2$ then it
is obvious that the previous reasoning is true and the properties
(\ref{eil}), (\ref{equon}) and (\ref{nieu}) hold for $J^2$.  However
we obtain new restrictions by requiring that the transformation
corresponding to $\varepsilon_1$ and $\varepsilon_2$ close to a
supersymmetry. The algebra \be\lb{algebra} [\delta(\varepsilon_1),
\delta(\varepsilon_2)]\varphi^A=2
\overline{\varepsilon}^i_2\gamma^{\mu}\varepsilon^j_1 \delta_{ij}
\partial_{\mu}\varphi^A
\ee (which correspond to (\ref{tecnic}) with $S^A_{BC}=0$) is
obtained if and only if \be\lb{cuatro}
 J^i\cdot J^j + J^j\cdot J^i = \delta_{ij}I
\ee \be\lb{cruzado} N^{ij}(X,Y)=[X,Y]+J^{i}[X,J^{j}Y]+J^{i}[J^{j}X,Y]-[J^{i} X,J^{j}Y]=0.
\ee
$N^{ij}(X,Y)$ is known as the mixed Niejenhuis tensor.

  Let us define the tensor $J^3=J^1 \cdot J^2$ for a $N=(3,0)$ supersymmetric
sigma model. It follows from (\ref{cuatro}) that
$$
(J^3)^2=-I
$$
and therefore $J^3$ is also an almost complex structure. One can
easily verify that 
\be\lb{lie} J^i\cdot J^j=-\delta_{ij}+ \epsilon^{ijk}J^k. \ee
 The integrability of
$J^1$ and $J^2$ implies the integrability of $J^3$ and
(\ref{cruzado}). Also (\ref{equon}) for $J^1$ and $J^2$ imply
that \be\lb{shushu} D_A(J^3)^B_C=0. \ee Therefore $N=(3,0)$
supersymmetry implies $N=(4,0)$ supersymmetry for sigma models and
from (\ref{cruzado}), (\ref{torso}), (\ref{equon}) and (\ref{lie})
it follows that the target metric $g_{AB}$ of a $N=(4, 0)$
supersymmetric sigma model is always \emph{strong hyperkahler
torsion}. If the form $b_{AB}$ is zero, then the geometry
presented here reduces to the usual Kahler and hyperkahler ones.

   The weak cases are also of interest in the context of low energy
heterotic string actions because they contain a three form that is
not closed due to a Green-Schwarz anomaly. For more details of
this assertions see \cite{Callan}-\cite{Stromin}.

\end{document}